\documentclass[12pt]{article}

\usepackage{amsfonts,amssymb,amsmath}

\usepackage{graphicx}
\DeclareGraphicsExtensions{.jpg,.pdf,.png}

\begin{document}
	
\title{\bf Refrigeration by modified Otto cycles and modified swaps through generalized measurements}
\author{ Naghi Behzadi $^{a}$ 
\\ $^a${\small Department of Theoretical Physics and Astrophysics,}
\\ {\small University of Tabriz 5166614776, Tabriz, Iran}}\maketitle

\begin{abstract}
We introduce two types of thermodynamic refrigeration cycles obtained through modification of the Otto cycle refrigerator by a generalized measurement channel. These refrigerators are corresponding to the activation of the measurement-based stroke before (first type) and after (second type) the full thermalization of the cooling medium by the cold reservoir in the related familiar Otto cycle. We show that the coefficient of performance for the first type modified refrigerator increases linearly in terms of measurement strength parameter, beyond the classical cooling of the familiar Otto cycle refrigerator. The second type interestingly introduces an autonomous refrigerator whose supplying work is provided by a quantum engine induced by the measurement channel and operates inside the modified cycle. By the considered measurement channel, we also establish such modifications on the swap refrigerator. It is observed that the thermodynamic properties of the obtained modified swap refrigerators are the same as of the modified Otto cycle ones respectively.         
\end{abstract}

\section{Introduction}
The main landscape of quantum thermodynamics is concerned with how the effect of quantum resources such as coherence, entanglement and quantum measurement on the quality of work extraction, refrigeration, etc, in thermal devices can be understood and applied [1-5]. As is well-known, quantum measurements are able to alter the average energy and entropy of a quantum system, provided that the measured observable does not commute with the Hamiltonian of that system [6-10]. Thus they can play similar role as thermal reservoirs. The measurement-driven devices which use energetic fluctuations generated by quantum measurement, have recently been explored [5, 9-14].  

On the other hand, in Ref. \cite{Behzadi}, a three-stroke quantum engine on the basis of non-selective generalized measurement channels with adjustable measurement strengths has been proposed and later, by employing nuclear magnetic resonance (NMR) platform, it has been realized experimentally \cite{Lisboa}. An interesting advantage of the introduced engine in \cite{Behzadi} which was also observed in the experimental results reported in \cite{Lisboa}, is that this kind of quantum engine can reach unit efficiency while also achieving maximum extracted power at the same time with the fine-tuning of the measurement strengths parameter. The approach of \cite{Behzadi} then was developed to introduce another measurement-based devices such as refrigerator and thermal accelerator according to a measurement setting intensity \cite{Dieguez}. 

The other feature of measurement channels introduced in \cite{Behzadi} is that their implementation running time is much smaller than any decoherence or thermalization time scales \cite{Lisboa}. This property along with the adjustable measurement strength, encourage one to involve this type of measurement channel, as a quantum resource, in the known thermodynamics cycles such as Otto cycle and swap, as an additional stroke. Involving the thermodynamics cycles with this additional stroke, called as modified cycles, may lead the obtained devices to have surpassing properties relative to the traditional ones. Strictly speaking, we note that some measurement-fueled quantum machines have been proposed with potential to approach beyond the Carnot efficiency, which are strictly related to the use of measurement process selectively (in the presence of feedback control) \cite{Elouard2,Elouard3,Jordan} or non-selectively \cite{Behzadi,Dieguez}. However, it is useful to investigate the advantages of involving a non-equilibrium energy resource provided by a measurement apparatus, into equilibrium thermodynamic cycles with, e.g., Otto efficiency. It would be interesting when the efficiency of such modified cycles go beyond the Otto limit.  

In this work, we perform two types of modifications on the familiar Otto cycle using a generalized measurement channel similar to the one introduced in \cite{Behzadi}, as an additional stroke. As is well-known in the Otto cycle, a quantum system, particularly a two-level one considered as cooling medium (CM), undergoes two isochoric and two adiabatic strokes. The modifications on the Otto cycle refrigerator are obtained whenever the measurement-based stroke is exploited before (first type) and after (second type) the full thermalization of the CM by the cold thermal reservoir. The first type modification leads to a refrigerator whose coefficient of performance (COP) increases in terms of measurement strength parameter beyond the case related to the Otto cycle refrigerator, which is a genuine quantum effect. In fact, for a given value of external work, the COP for modified Otto cycle refrigerator goes beyond the familiar Otto cycle one depending on the intensity of measurement-based stroke. 

The appearance of such feature in the first type modified refrigerator is due to involvement of non-equilibrium energy resource into the familiar Otto cycle (as a cycle of equilibrium thermodynamic operations) provided by a measurement apparatus. Recently in Ref. \cite{Yi}, it was introduced a single-temperature heat engine with a two-level working substance, similar to the Otto cycle one with difference that one of the thermalization strokes being replaced by a non-selective projective measurement. It was, however, observed that the efficiency doesn't exceed from the Otto efficiency related to the fact that the measurement apparatus affects the working substance similar to an equilibrium thermal reservoir with infinite temperature. Such an Otto cycle was also examined using more complex working substances such as qutrit and two-qubit systems \cite{Anka}. The efficiencies of the obtained engines can exceed the Otto limit whenever there exist idle level(s), i.e. level(s) does (do) not couple to the work source. Also, based on Ref. \cite{Yi}, a measurement-based quantum heat engine with interacting two spin-$\frac{1}{2}$s, two spin-$1$s and two spin-$\frac{3}{2}$s as working substances, was introduced in \cite{Das} such that the effects of coupling strength of the spins along with the dimension of corresponding Hilbert space on the efficiency were investigated. In this direction, a finite time version of single temperature heat engine \cite{Yi}, with interacting two-spin working substance, was developed in \cite{Purkait}. For this engine, the measurement-based stroke is provided by global non-selective projective measurement in the Bell basis. By adapting optimal choice for finite time tunning of unitary stages, the efficiency of this engine approaches to the Otto limit. Another measurement-based single-temperature thermal machine in which the working substance is a single-mode quantum harmonic oscillator introduced in \cite{Santos}. In this engine, two non-unitary parts are provided by thermal reservoir structured with $\mathcal{PT}$-symmetric non-Hermitian Hamiltonians and a position measurement process such that the cycle can operate in the engine and refrigerator modes only in the Otto limit. Also, in a different situation, a quantum heat engine was formulated in \cite{Hayashi} where measurement process is employed as a completely positive work extraction instrument. The impact of different measurement-based diagnostic methods on the efficiency and output power of a quantum Otto engine with two-level working substance, as well as on the charging process of a quantum battery provided by a quantum Otto machine were analyzed in Refs. \cite{Son,Sonn} respectively.

The second type modification interestingly leads to another type of quantum autonomous refrigerator which by itself have been subject of researches in last years [25-36]. In general, a refrigerator is autonomous in the sense that it does not require any external work sources for cooling. This process is a quantum effect because the required work for the refrigeration is supplied by a three-stroke engine whose strokes are directly dependent on the measurement-based stroke. It is interesting to note that the efficiency of this engine is the same as of an Otto engine. 

The modification scheme described for the Otto cycle refrigerator can be applied easily for the two-stroke swap refrigerator too. As the previous case, we can obtain two types of three-stroke modified swap refrigerators depending on that the measurement-based stroke activated before (first type) or after (second type) the swap operation. The obtained modified swap refrigerators have the same properties as of the modified Otto cycle refrigerators respectively.  

\section{Refrigeration by first type modified Otto cycle} 
In a four-stroke Otto cycle refrigerator, investing an amount of work supplied by an external agent on the CM (during adiabatic strokes) causes remove of an amount of heat from the cold reservoir and loss of another amount of heat into the hot reservoir (during two isochoric strokes). In this section, by considering the familiar Otto cycle in the refrigeration mode, we involve a measurement apparatus into the cycle before full thermalization of the CM by the cold reservoir. We show that the obtained modified refrigerator removes more heat from the cold reservoir for the same amount of invested work. To this aim, let us consider that the CM is a two-level system with Hamiltonian $H=\frac{\omega}{2}\sigma_{z}$ where $\sigma_{z}$ is the z-component Pauli spin-$\frac{1}{2}$ operator and $\omega$ is the corresponding transition frequency. In general, a measurement channel makes generalized measurements on the CM described by completely positive and trace-preserving (CPTP) maps \cite{Jacobs}. To describe the first type modified Otto cycle, let us introduce a particular measurement on the CM represented by the following measurement operators with adjustable measurement strength \cite{Behzadi}, as follows 

\begin{eqnarray}
M_{1}(\mathcal{\xi})=|0\rangle\langle 0|+\sqrt{1-\mathcal{\xi}}|1\rangle\langle 1|, \qquad M_{2}(\mathcal{\xi})=\sqrt{\mathcal{\xi}}|0\rangle\langle 1|,
\end{eqnarray}    
where $\mathcal{\xi}$ is the measurement strength parameter.

Obviously the operators $M_{1}(\mathcal{\xi})$ and $M_{2}(\mathcal{\xi})$ satisfy $\Sigma_{i}M_{i}(\mathcal{\xi})^{\dagger}M_{i}(\mathcal{\xi})=I$ which is, in fact, the essence of fundamental theorem of quantum measurement \cite{Jacobs}. In addition, the operators $E_{1}(\xi):=M_{1}(\mathcal{\xi})^{\dagger}M_{1}(\mathcal{\xi})=|0\rangle\langle 0|+\left(1-\mathcal{\xi}\right)|1\rangle\langle 1|$ and $E_{2}(\xi):=M_{2}(\mathcal{\xi})^{\dagger}M_{2}(\mathcal{\xi})=\mathcal{\xi}|1\rangle\langle 1|$ are the so-called POVM (positive operator valued measure) operators corresponding to the measurement operators (1). Clearly, $0\le\mathcal{\xi}\le1$, where $\mathcal{\xi}=0$ indicates that the system has not been disturbed (measured) by the measurement process, and $\mathcal{\xi}=1$ corresponds to performing strong or projective measurement on the CM.

To achieve first modification on the Otto cycle (Fig. 1), let us assume at the beginning that the CM is in thermal equilibrium with the cold reservoir whose inverse temperature is $\beta_{c}$. In this time, the Hamiltonian of the CM is denoted by $H_{c}=\frac{\omega_{c}}{2}\sigma_{z}$ where $\omega_{c}$ is the corresponding transition frequency. So the corresponding Gibbs state of the CM is denoted by $\rho_{c}=e^{-\beta_{c}H_{c}}/\mathcal{Z}_{c}$, where $\mathcal{Z}_{c}=Tr\left(e^{-\beta_{c}H_{c}}\right)$ is the partition function. The mean energy, in this step, is equal to $E_{1}=Tr\left(H_{c}\rho_{c}\right)=-\frac{\omega_{c}}{2}\mathrm{tanh}\left(\beta_{c}\omega_{c}/2\right)$. To describe the 1st stroke, we let the CM is isolated from the cold reservoir and undergoes adiabatic unitary evolution in which only the transition frequency (for simplicity and without lose of generality) changes from $\omega_{c}$ to $\omega_{h}$ such that $\omega_{c}<\omega_{h}$. Therefore, the mean energy after this stroke is obtained as $E_{2}=Tr\left(H_{h}\rho_{c}\right)=-\frac{\omega_{h}}{2}\mathrm{tanh}\left(\beta_{c}\omega_{c}/2\right)$ where $H_{h}=\frac{\omega_{h}}{2}\sigma_{z}$.

The 2nd stroke is provided through full thermalization of the CM by a hot thermal reservoir with inverse temperature $\beta_{h}$ $(\beta_{h}<\beta_{c})$. In this time the corresponding Gibbs state is given by $\rho_{h}=e^{-\beta_{h}H_{h}}/\mathcal{Z}_{h}$ where $\mathcal{Z}_{h}=Tr\left(e^{-\beta_{h}H_{h}}\right)$ is the partition function. By this stroke, the mean energy is changed and is equal to  $E_{3}=Tr\left(H_{h}\rho_{h}\right)=-\frac{\omega_{h}}{2}\mathrm{tanh}\left(\beta_{h}\omega_{h}/2\right)$. In the 3rd stroke, which is similar to the first one, the CM is isolated from the hot reservoir and undergoes adiabatic evolution in which the transition frequency is brought into the original value $\omega_{c}$. The corresponding mean energy becomes as  $E_{4}=Tr\left(H_{c}\rho_{h}\right)=-\frac{\omega_{c}}{2}\mathrm{tanh}\left(\beta_{h}\omega_{h}/2\right)$.

The promised modification for the familiar Otto cycle is introduced by the 4th stroke where the CM is let to interact with a measurement apparatus described by non-selective generalized measurement as $\mathcal{E}(\rho_{h})=\Sigma_{i}M_{i}(\xi)\rho_{h}M_{i}^{\dagger}(\xi)$ such that $M_{i}(\xi)$s are the measurement operators defined by Eq. (1). After this stroke, the mean energy is 
\begin{eqnarray}
E_{5}=Tr\left(H_{c}\rho_{_{h}}(\mathcal{\xi})\right)=-\frac{\omega_{c}}{2}\mathrm{tanh}\left(\beta_{h}\omega_{h}/2\right)-\mathcal{\xi}\omega_{c}\frac{e^{-\beta_{h}\omega_{h}/2}}{\mathcal{Z}_{h}},
\end{eqnarray} 
where 
\begin{eqnarray}
\mathcal{E}(\rho_{h}):=\rho_{_{h}}(\mathcal{\xi})=\frac{1}{\mathcal{Z}_{h}}\left((1-\mathcal{\xi})e^{{-\beta_{h}\omega_{h}/2}}|1\rangle\langle 1|+(e^{{\beta_{h}\omega_{h}/2}}+\mathcal{\xi}e^{{-\beta_{h}\omega_{h}/2}})|0\rangle\langle 0|\right).
\end{eqnarray}
It is clear that $\rho_{_{h}}(\mathcal{\xi})$ is a non-equilibrium passive state with no induced ergotropy \cite{Allahverdyan}. Consequently, the cycle is completed by 5th stroke which is the thermalization of the CM by the cold reservoir. By considering the condition $\beta_{c}\omega_{c}<\beta_{h}\omega_{h}$ (refrigerator mode), the CM throughout the modified Otto cycle exchanges following works and heats with work and heat sources as well as with the measurement apparatus as 
\begin{eqnarray}
1\mathrm{st} \quad \mathrm{stroke}: \quad W_{1}=E_{2}-E_{1}=\left(\frac{\omega_{c}-\omega_{h}}{2}\right)\mathrm{tanh}\left(\frac{\beta_{c}\omega_{c}}{2}\right),
\end{eqnarray}
\begin{eqnarray}
2\mathrm{nd} \quad \mathrm{stroke}: \quad
Q_{h}=E_{3}-E_{2}=\frac{\omega_{h}}{2}\left(\mathrm{tanh}\left(\frac{\beta_{c}\omega_{c}}{2}\right)-\mathrm{tanh}\left(\frac{\beta_{h}\omega_{h}}{2}\right)\right),
\end{eqnarray} 
\begin{eqnarray}
3\mathrm{rd} \quad \mathrm{stroke}: \quad
W_{2}=E_{4}-E_{3}=\left(\frac{\omega_{h}-\omega_{c}}{2}\right)\mathrm{tanh}\left(\frac{\beta_{h}\omega_{h}}{2}\right),
\end{eqnarray}
\begin{eqnarray}
4\mathrm{th} \quad \mathrm{stroke}: \quad
Q(\xi)=E_{5}-E_{4}=-\mathcal{\xi}\omega_{c}\frac{e^{-\beta_{h}\omega_{h}/2}}{\mathcal{Z}_{h}},
\end{eqnarray}
\begin{eqnarray}
5\mathrm{th} \quad \mathrm{stroke}: \quad
Q_{c}(\xi)=E_{1}-E_{5}=Q_{c}-Q(\xi),
\end{eqnarray} 
where
\begin{eqnarray}
Q_{c}=\frac{\omega_{c}}{2}\left(\mathrm{tanh}\left(\frac{\beta_{h}\omega_{h}}{2}\right)-\mathrm{tanh}\left(\frac{\beta_{c}\omega_{c}}{2}\right)\right).
\end{eqnarray} 
$Q_{h}<0$ ($Q(\xi)<0$) is the amount of heat (``quantum heat'' \cite{Buffoni,Behzadi}) flowed from the CM into the hot reservoir (measurement apparatus) while $Q_{c}(\xi)>0$ is as heat flowed from the cold reservoir into the CM (Fig. 1). Also the work $W_{1}<0$ is extracted from the CM while $W_{2}>0$ is invested on it as depicted in Fig. 1. Clearly in the absence of measurement process $(\xi=0)$, the removed heat from the cold reservoir is reduced to $Q_{c}$, corresponding to the case of Otto cycle refrigerator. Hence the COP for the modified refrigerator is obtained as follows
\begin{eqnarray}
\mathrm{COP}(\xi)=\frac{Q_{c}(\xi)}{W_{in}}=\left(\frac{\omega_{h}}{\omega_{c}}-1\right)^{-1}\left(1+\frac{2\mathcal{\xi}}{\mathcal{Z}_{h}}\frac{e^{-\beta_{h}\omega_{h}/2}}{\left(\mathrm{tanh}\left(\frac{\beta_{h}\omega_{h}}{2}\right)-\mathrm{tanh}\left(\frac{\beta_{c}\omega_{c}}{2}\right)\right)}\right),
\end{eqnarray} 
where
\begin{eqnarray}
W_{\mathrm{in}}=W_{1}+W_{2}=\left(\frac{\omega_{h}-\omega_{c}}{2}\right)\left(\mathrm{tanh}\left(\frac{\beta_{h}\omega_{h}}{2}\right)-\mathrm{tanh}\left(\frac{\beta_{c}\omega_{c}}{2}\right)\right),
\end{eqnarray}
is the required work $(W_{\mathrm{in}}>0)$ for the refrigeration which should be invested on the CM by an external agent to remove heat from the cold reservoir. It is clear that when the measurement strength parameter becomes zero, i.e $\mathcal{\xi}=0$, then the COP in Eq. (10) becomes the same as of the familiar Otto cycle refrigerator. 

It is useful to make a comparison between the refrigeration scheme obtained through modified Otto cycle and that one investigated in \cite{Dieguez}, where the proposed scheme is based on a single temperature three-stroke cycle. To this aim, we consider that the net invested work in Eq. (11), in fact, depends on the two equilibrium work strokes as well as on (perfect) thermalization with the hot reservoir. But in the three-stroke cycle, the invested work is related only to a non-equilibrium state obtained by a measurement channel in an isentropic way (work stroke). Roughly speaking, the mentioned three strokes of the first type modified Otto cycle is equivalent to the first (work) stroke of the three-stroke cycle. Obviously, the remainder strokes are the same for each cycle. Hence it is natural to adopt the same definition for COP for each cycle.     

Fig. 2, shows the COP for the modified refrigerator in terms of measurement strength parameter $\xi$, for various amounts of invested work. In fact, for each values of invested works, the COP is always a linearly increasing function of $\xi$ and goes beyond the familiar Otto cycle case. It is appear that the rate of its increment is inversely proportional to the invested works. Therefore, the presence of generalized measurement (1) in the Otto cycle, as a quantum resource, leads to the supremacy of  COP for the (first type) modified refrigerator with respect to the classical Otto cycle refrigerator. On the other hand, the cooling process characterized by the amount of removal heat from the cold reservoir, i.e. $Q_{c}(\xi)$ in Eq. (8), is also an increasing function of $\xi$ (Fig. 3). However, it is interesting to note that as $\xi\rightarrow1$, corresponding to the strong measurement, the cooling process approaches to a unique value irrespective to the amount of invested works. In fact, in the strong measurement ($\xi=1$) the CM is found in its ground state, i.e. $E_{5}=-\frac{\omega_{c}}{2}$ as well as its von-Neumann entropy becomes zero. So the removal heat is $Q_{c}(\xi)=\frac{\omega_{c}}{2}\left(1-\mathrm{tanh}({\beta_{c}\omega_{c}}/2)\right)$ which is clearly the same for each amount of invested works.  

We should note that the generalized measurement defined by Eq. (1), has been involved in the Otto cycle refrigerator non-selectively, i.e. without post-selection of measurement outcomes. Similar situations are observed in \cite{Yi,Behzadi,Dieguez} such that non-selective projective measurement plays the role of hot thermal reservoir in \cite{Yi} and non-selective generalized measurements with adjustable measurement strength are responsible for generation of quantum heat and extraction of work in the considered thermodynamic cycles \cite{Behzadi,Dieguez}. In quantum Maxwell's demon engines [11,12,40-42], work is extracted by assisting a feedback control based on performed measurement by a demon. It is indeed treated as a device extracting information from the working substance by measurement and storing it onto some classical memory (readout), and exerting some action on working substance, conditioned to the readout (controlled or selective measurement) \cite{Elouard2}. At the end of each cycle the memory of the demon is erased by some nonzero work cost. Obviously such a work cost is not taken into account for the cases in which measurement process is exploited non-selectively only as a energy resource as well as for the case of introduced refrigerator in this work.

We conclude that the requirement for use of feedback from a measurement result to extract work depends on the property of corresponding thermodynamic cycle. For example, in quantum Szilard engine \cite{Kim}, extracting work from a thermal reservoir is possible only by utilizing information acquired by a demon. However, in quantum Stirling engine whose working substance is the same as of the quantum Szilard engine, work is extracted without requirement to perform measurement and feedback of its result \cite{Thomas, Dass} (see also \cite{Chattopadhyay1,Chattopadhyay2}), i.e. the presence of a demon is not necessary here as well as for the discussed modified refrigerator. 

The other point which should be properly investigated here is related to the term ``quantum heat'' which we used in this section (Eq. 7) and use it again in the next sections. It was introduced in \cite{Elouard1}, as consequence of quantum measurement and later used in [10-12,15-17]. It is obvious that measuring a quantum system in a basis that does not commute with its Hamiltonian leads energy to be taken away from the measurement apparatus and given to the system and vice versa. The exchanged energy is stochastic in nature and so has some similarities to heat in a stochastic thermodynamics context \cite{Elouard1}. However, this similarity is only superficial, because in measurement fueled quantum engines the Carnot bound (or other existing thermodynamic bounds) may not apply, and these engines (by well-chosen measurement process) are able to convert the given “quantum heat” to useful work with efficiency approaches to unity. In other words, “quantum heat” may have a random or ordered nature depending on measurement device shows behaviors somewhere in between a thermal reservoir and a work source. Finally, it is emphasized that the obtained results in this section, i.e. Figs. 2, 3, (as well as of the next sections) are merely consequences of involving quantum measurement into Otto cycle and they are not arisen from other quantum resources such as quantum coherence which in turns can be exploited as a fuel in some quantum machine protocols [50-52].     

\section{Refrigeration by second type modified Otto cycle}
Let us consider the other type of modified Otto cycle obtained by activation of the measurement channel after the full thermalization of the CM by the cold reservoir, as shown in Fig. 4. The 1st stroke causes flow of quantum heat from the CM into the measurement apparatus calculated as follows
\begin{eqnarray}
Q(\xi)=E_{2}-E_{1}=-\xi\omega_{c}\frac{e^{-\beta_{c}\omega_{c}/2}}{\mathcal{Z}_{c}}<0,
\end{eqnarray}
where $E_{1}=Tr\left(H_{c}\rho_{c}\right)$, as seen in the previous section, is the mean energy at the beginning of the 1st stroke and  
\begin{eqnarray}
E_{2}=Tr\left(H_{c}\rho_{c}(\xi)\right)=-\frac{\omega_{c}}{2}\mathrm{tanh}\left(\frac{\beta_{c}\omega_{c}}{2}\right)-\xi\omega_{c}\frac{e^{-\beta_{c}\omega_{c}/2}}{\mathcal{Z}_{c}},
\end{eqnarray} 
is at the end of it with 
\begin{eqnarray}
\rho_{c}(\xi)=\frac{1}{\mathcal{Z}_{c}}\left((1-\xi)e^{{-\beta_{c}\omega_{c}/2}}|1\rangle\langle 1|+(e^{{\beta_{c}\omega_{c}/2}}+\xi e^{{-\beta_{c}\omega_{c}/2}})|0\rangle\langle 0|\right).
\end{eqnarray}
Obviously the non-equilibrium state (14), is a passive state without ergotropy. By the 2nd stroke, the CM evolves adiabatically such that only its transition frequency is changed from $\omega_{c}$ to $\omega_{h}$ $(\omega_{c}<\omega_{h})$. Therefore, the mean energy at the end of this stroke becomes 
\begin{eqnarray}
E_{3}=Tr\left(H_{h}\rho_{c}(\xi)\right)=-\frac{\omega_{h}}{2}\mathrm{tanh}\left(\frac{\beta_{c}\omega_{c}}{2}\right)-\mathcal{\xi}\omega_{h}\frac{e^{-\beta_{c}\omega_{c}/2}}{\mathcal{Z}_{c}},
\end{eqnarray} 
where  $H_{h}=\frac{\omega_{h}}{2}\sigma_{z}$. The extracted work from the CM through this stroke is written as 
\begin{eqnarray}
\mathcal{W}(\xi)=E_{3}-E_{2}=W_{1}+W(\xi),
\end{eqnarray}
where 
\begin{eqnarray}
W_{1}=\left(\frac{\omega_{c}-\omega_{h}}{2}\right)\mathrm{tanh}\left(\frac{\beta_{c}\omega_{c}}{2}\right),
\end{eqnarray}
and
\begin{eqnarray}
W(\xi)=\mathcal{\xi}(\omega_{c}-\omega_{h})\frac{e^{-\beta_{c}\omega_{c}/2}}{\mathcal{Z}_{c}}.
\end{eqnarray}
$\mathcal{W}(\xi)$ is always negative because $W_{1}$ and $W(\xi)$ are always negative. The expression (18), as a part of the extracted work in Eq. (16), originates only from the measurement process such that  $\mathcal{W}(\xi=0)=W_{1}$. On the other hand, during the 3rd stroke, the CM is brought into contact with the hot thermal reservoir so that the amount of exchanged heat at the end of this stroke becomes as below
\begin{eqnarray}
Q_{h}(\xi)=E_{4}-E_{3}=Q_{h}+Q'(\xi),
\end{eqnarray} 
where
\begin{eqnarray}
Q_{h}=\frac{\omega_{h}}{2}\left(\mathrm{tanh}\left(\frac{\beta_{c}\omega_{c}}{2}\right)-\mathrm{tanh}\left(\frac{\beta_{h}\omega_{h}}{2}\right)\right),
\end{eqnarray} 
\begin{eqnarray}
Q'(\xi)=\xi\omega_{h}\frac{e^{-\beta_{c}\omega_{c}/2}}{\mathcal{Z}_{c}}>0,
\end{eqnarray} 
and $E_{4}=Tr\left(H_{h}\rho_{h}\right)$. By considering the refrigeration condition, i.e. $\beta_{c}\omega_{c}<\beta_{h}\omega_{h}$, we have $Q_{h}<0$. Thus by this stroke, there is flow of heat $Q_{h}$ from the CM into the hot reservoir and at the same time, flow of another heat $Q'(\xi)$ from the hot reservoir into the CM, which is only originated from the effect of quantum measurement in the 1st stroke. In the 4th stroke, the Hamiltonian of the CM is returned to its original form, i.e. only its transition frequency is changed from $\omega_{h}$ to $\omega_{c}$ adiabatically. The invested work, by this stroke, on the CM is calculated as 
\begin{eqnarray}
W_{2}=E_{5}-E_{4}=\left(\frac{\omega_{h}-\omega_{c}}{2}\right)\mathrm{tanh}\left(\frac{\beta_{h}\omega_{h}}{2}\right),
\end{eqnarray}
where $W_{2}>0$ and $E_{5}=Tr\left(H_{c}\rho_{h}\right)$. Finally, the CM is brought again into contact with the cold reservoir and the cycle is completed by this (5th) stroke so the amount of heat flowed from the cold reservoir into the CM is obtained as follows
\begin{eqnarray}
Q_{c}=E_{1}-E_{5}=\frac{\omega_{c}}{2}\left(\mathrm{tanh}\left(\frac{\beta_{h}\omega_{h}}{2}\right)-\mathrm{tanh}\left(\frac{\beta_{c}\omega_{c}}{2}\right)\right),
\end{eqnarray} 
where $Q_{c}>0$. The net work which should be invested by an external agent on the CM throughout the modified cycle is
\begin{eqnarray}
W_{ex}(\xi)=W_{in}+W(\xi)=\left(\frac{\omega_{h}-\omega_{c}}{2}\right)\left(\mathrm{tanh}\left(\frac{\beta_{h}\omega_{h}}{2}\right)-\mathrm{tanh}\left(\frac{\beta_{c}\omega_{c}}{2}\right)-2\xi\frac{e^{-\beta_{c}\omega_{c}/2}}{\mathcal{Z}_{c}}\right),
\end{eqnarray}
where $W_{ex}(\xi)\ge0$. On the other hand, $W_{in}=-(Q_{h}+Q_{c})=W_{ex}(\xi)-W(\xi)$ is the amount of work required to invest on the CM in order to remove the heat $Q_{c}$ from the cold reservoir and transfer the heat $Q_{h}$ into the hot reservoir. It is explicitly independent from the measurement-based stroke whereas $W_{ex}(\xi)$ and $W(\xi)$ are not. The COP for the refrigeration according to second type modified Otto cycle is 
\begin{eqnarray}
\mathrm{COP}(\xi)=\frac{Q_{c}}{W_{ex}(\xi)-W(\xi)},
\end{eqnarray}
which is also independent from the measurement-based stroke too, and as is evident, it is the same as COP of familiar Otto cycle refrigerator. So what does measurement do really? To answer to this question, let us at first consider the case that the measurement process (1) has not been activated, i.e. $\xi=0$. In this case $W(\xi=0)=0$ so $W_{in}=W_{ex}(\xi=0)$, thus the required work for the refrigeration is provided only by an external agent. 

Now, in the presence of measurement-based stroke, i.e. $\xi\ne0$, a three-stroke measurement-induced quantum engine is activated such that whose operation cycle is along the modified cycle. This cycle consists of absorption of heat $Q'(\xi)$ from the hot reservoir and transfer a part of it, i.e. $Q(\xi)$, into the measurement apparatus along with extraction of work $W(\xi)$ such that $W(\xi)=-(Q'(\xi)+Q(\xi))$, as shown in Fig. 4. Strictly speaking, along the modified cycle, two machines works simultaneously one as a familiar Otto cycle refrigerator independent from the measurement process and the other as an engine whose efficiency is the same as of an Otto engine and strictly depends on the measurement (see Fig. 4). Hence, the extracted work can be regarded (by a suitable feedback) as a part of required work for the refrigeration so the remainder part is supplied by an external agent ($W_{in}=W_{ex}(\xi)-W(\xi)$). Furthermore, since $W_{ex}(\xi)$ is a linearly decreasing function of $\xi$, then it becomes zero for the following value of measurement strength parameter as
\begin{eqnarray}
\xi_{c}=\mathcal{Z}_{c}\frac{e^{\beta_{c}\omega_{c}/2}}{2}\left[\mathrm{tanh}\left(\frac{\beta_{h}\omega_{h}}{2}\right)-\mathrm{tanh}\left(\frac{\beta_{c}\omega_{c}}{2}\right)\right],
\end{eqnarray} 
i.e. $W_{ex}(\xi_{c})=0$, thus $W_{in}=-W(\xi_{c})$ (note that $\xi_{c}<1$). 

The last relation explicitly shows that the required work for the refrigeration is only supplied by the measurement-induced engine. So it is natural to conclude that the second type modified Otto cycle presents an autonomous refrigerator whose required work for the refrigeration is provided only by the measurement-induced engine which acts simultaneously along the modified cycle. Trivially, for $\xi_{c}<\xi\le1$, the measurement-induced engine not only provides the required work for the refrigerator but also can gives the work $W_{ex}(\xi)<0$ into another work source.  

There are also other types of modification through participating the measurement based-stroke into the familiar Otto cycle other than the cases discussed  previously. In fact, these modifications are not effective because they do not lead to any new advantages as well as remarkable improvement on the outputs of obtained cycles in comparison to the familiar Otto cycle.  
 
\section{Refrigeration by modified swaps}
In the previous sections, we obtained two types of refrigerators on the basis of modifying the Otto cycle refrigerator using the generalized measurement (1). In this section, we perform such modifications on the two-stroke swap refrigerator. We see that the obtained refrigeration advantages for the modified swaps are the same as of the modified Otto cycle. Let us consider two two-level systems with Hamiltonians $H_{i}=\frac{\omega_{i}}{2}\sigma_{z}^{i}$ ($i=h,c$) with $\omega_{c}<\omega_{h}$, each of which is in thermal contact with thermal reservoirs whose inverse temperatures are $\beta_{h}$ and $\beta_{c}$ $(\beta_{h}<\beta_{c})$ respectively. So in the thermal equilibrium, the corresponding Gibbs states are $\rho_{h}$ and $\rho_{c}$ respectively (see Sec. II). 

To describe the first type modified swap (Fig. 5), consider that each subsystem has been isolated from its own reservoir such that they are described by a product state as $\rho=\rho_{h}\otimes\rho_{c}$. Now consider that this bipartite system is exchanged by the following swap operator
\begin{eqnarray}
U=|11\rangle\langle11|+|10\rangle\langle01|+|01\rangle\langle10|+|00\rangle\langle00|,
\end{eqnarray} 
which is an unitary operator. Since the entropy of the bipartite system is left unchanged under this operation so the amount of change in mean energy is interpreted as work as follows  
\begin{eqnarray}
\begin{split}
&W_{in}=Tr\left(HU\rho U^{\dagger}\right)-Tr\left(H\rho\right)\\&=\left(\frac{\omega_{h}-\omega_{c}}{2}\right)\left(\mathrm{tanh}\left(\frac{\beta_{h}\omega_{h}}{2}\right)-\mathrm{tanh}\left(\frac{\beta_{c}\omega_{c}}{2}\right)\right).
\end{split}
\end{eqnarray}
where $H=H_{h}+H_{c}$ and $U\rho U^{\dagger}=\rho_{c}\otimes \rho_{h}$. The invested work in Eq. (28), is maximal \cite{Campisi}, which is the same as of Eq. (11) in Sec. II. In the 2nd stroke, the subsystem with Hamiltonian $H_{c}$ and density operator $\rho_{h}$ is affected by the generalized measurement (1), so the state of the whole system after the measurement is left as $(I\otimes \mathcal{E})U\rho U^{\dagger}=\rho_{c}\otimes\mathcal{E}(\rho_{h})$ where $\mathcal{E}(\rho_{h})=\Sigma_{i}M_{i}(\xi)\rho_{h}M_{i}^{\dagger}(\xi)$ and $M_{i}(\xi)$s are the measurement operators defined in Eq. (1). By this stroke, the amount of quantum heat flowed from this subsystem into the measurement apparatus is 
\begin{eqnarray}
Q(\xi)=Tr(H\rho_{c}\otimes\mathcal{E}(\rho_{h}))-Tr(H\rho_{c}\otimes \rho_{h})=-\mathcal{\xi}\omega_{c}\frac{e^{-\beta_{h}\omega_{h}/2}}{\mathcal{Z}_{h}},
\end{eqnarray}
which is the same as Eq. (7), in Sec. II. The 3rd stroke is provided by returning each subsystem into contact with its own reservoir and the cycle (modified swap) is completed. Hence the exchanged heat between each subsystem and its respective reservoir is obtained as below   

\begin{eqnarray}
Q_{h}=Tr(H_{h}\rho_{h})-Tr(H_{h}\rho_{c})=\frac{\omega_{h}}{2}\left(\mathrm{tanh}\left(\frac{\beta_{c}\omega_{c}}{2}\right)-\mathrm{tanh}\left(\frac{\beta_{h}\omega_{h}}{2}\right)\right), 
\end{eqnarray}
and 
\begin{eqnarray}
Q_{c}(\xi)=Tr(H_{c}\rho_{c})-Tr(H_{c}\mathcal{E}(\rho_{h}))=Q_{c}-Q(\xi),
\end{eqnarray}
where $Q_{h}$, $Q(\xi)$ and $Q_{c}$ are the same as of Eqs. (5), (7) and (9) in Sec. II, respectively. It is noted that in the refrigerator mode ($\beta_{c}\omega_{c}<\beta_{h}\omega_{h}$) we have $W_{in}>0$, $Q_{h}<0$ and $Q_{c}>0$. By this results, the corresponding COP becomes as
\begin{eqnarray}
\mathrm{COP}(\xi)=\left(\frac{\omega_{h}}{\omega_{c}}-1\right)^{-1}\left(1+\frac{2\mathcal{\xi}}{\mathcal{Z}_{h}}\frac{e^{-\beta_{h}\omega_{h}/2}}{\left(\mathrm{tanh}\left(\frac{\beta_{h}\omega_{h}}{2}\right)-\mathrm{tanh}\left(\frac{\beta_{c}\omega_{c}}{2}\right)\right)}\right),
\end{eqnarray}
which is the same as of Eq. (10), so the COP for first type modified swap is the same as of the first type modified Otto cycle.  

At the end, let us introduce the second type modified swap refrigerator. Fig. 6, shows the details of second type modified swap refrigerator. As the previous case, it is assumed that at the beginning of the 1st stroke, each subsystem has been isolated from its own reservoir so that they are described by the tensor product $\rho=\rho_{h}\otimes\rho_{c}$. In the 1st stroke, the subsystem which was in the thermal contact with the cold reservoir is affected by the measurement process (1) so that the state of the whole system becomes $\rho'=\rho_{h}\otimes \mathcal{E}(\rho_{c})$. The amount of quantum heat extracted from the considered subsystem by the measurement apparatus is   

\begin{eqnarray}
Q(\xi)=Tr\left(H\rho'\right)-Tr\left(H\rho\right)=-\mathcal{\xi}\omega_{c}\frac{e^{-\beta_{c}\omega_{c}/2}}{\mathcal{Z}_{c}},
\end{eqnarray}
which is the same as of Eq. (12), in Sec. III. The 2nd stroke is provided by exploiting the swap operator (27) corresponding to the work stroke. The amount of work which should be invested by an external agent in order to have refrigeration can be regarded as 
\begin{eqnarray}
\begin{split}
&W_{ex}(\xi)=Tr\left(HU\rho'U^{\dagger}\right)-Tr\left(H\rho'\right)\\&=W_{in}+W(\xi)=\left(\frac{\omega_{h}-\omega_{c}}{2}\right)\left(\mathrm{tanh}\left(\frac{\beta_{h}\omega_{h}}{2}\right)-\mathrm{tanh}\left(\frac{\beta_{c}\omega_{c}}{2}\right)-2\mathcal{\xi}\frac{e^{-\beta_{c}\omega_{c}/2}}{\mathcal{Z}_{c}}\right),
\end{split}
\end{eqnarray}
which is the same as of Eq. (24) in Sec. III. As discussed in the previous section, $W_{ex}(\xi)\ge0$, $W(\xi)\le0$ depend only on the measurement strength parameter in the 1st stroke and $W_{in}(\xi)=W_{ex}(\xi)-W(\xi)$ is the amount of required work which should be invested on the CM in order to have refrigeration after completion of each cycle. In the 3rd stroke, each subsystem is returned to contact with its own reservoir in order to recovering the respective original state of each subsystem so the cycle is completed. The heat exchanged between each subsystem and the respective thermal reservoir is obtained easily as
\begin{eqnarray}
\begin{split}
&Q_{h}(\xi)=Tr\left(H_{h}\rho_{h}\right)-Tr\left(H_{h}\rho''_{h}\right)\\&=Q_{h}+Q'(\xi)=\frac{\omega_{h}}{2}\left(\mathrm{tanh}\left(\frac{\beta_{c}\omega_{c}}{2}\right)-\mathrm{tanh}\left(\frac{\beta_{h}\omega_{h}}{2}\right)+2\mathcal{\xi}\frac{e^{-\beta_{c}\omega_{c}/2}}{\mathcal{Z}_{c}}\right),
\end{split}
\end{eqnarray}
and 
\begin{eqnarray}
\begin{split}
&Q_{c}=Tr\left(H_{c}\rho_{c}\right)-Tr\left(H_{c}\rho''_{c}\right)\\&=\frac{\omega_{c}}{2}\left(\mathrm{tanh}\left(\frac{\beta_{h}\omega_{h}}{2}\right)-\mathrm{tanh}\left(\frac{\beta_{c}\omega_{c}}{2}\right)\right),
\end{split}
\end{eqnarray}
where $\rho''_{h}=Tr_{c}(U\rho'U^{\dagger})$ and $\rho''_{c}=Tr_{h}(U\rho'U^{\dagger})$. Remarkably Eqs. (35) and (36) are the same as of Eqs. (19) and (23) in Sec III, respectively. In fact, $Q_{h}(\xi)$ is the heat exchanged between the subsystem with Hamiltonian $H_{h}$ and the hot reservoir while the $Q_{c}$ is the heat exchanged between the subsystem with Hamiltonian $H_{c}$ and the cold reservoir. 

Thus the second type modified swap, as the second type modified Otto cycle, removes heat $Q_{c}$ from the cold reservoir and transfer heat $Q_{h}$ into the hot reservoir by investing work $W_{in}=-(Q_{h}+Q_{c})$. As argumented in Sec. III, it is evident from Eq. (34) that this work is supplied by an external agent as well as by a three stroke measurement-induced engine, i.e. $W_{in}=W_{ex}(\xi)-W(\xi)$. The induced engine by the measurement, in turns, absorbs heat $Q'_{h}(\xi)$ from the hot reservoir and gives heat $Q_{h}(\xi)$ into the measurement apparatus and so its advantage is that the work $W(\xi)$ is extracted. Each stroke of this engine is measurement dependent. By using the results of the last section, in $\xi=\xi_{c}$, the refrigerator is only supplied by the measurement-induced engine (through suitable feedback) so that there is no need to anther work supplied by an external agent. Thus we obtain the other type of autonomous refrigerator constructed by the second type modified swap. It is again emphasized that along the modified swap, we have simultaneous actions of familiar swap refrigerator (with Otto COP) and three-stroke engine (with Otto efficiency), the first one is independent from the measurement process and the second one strictly depends of it. So by adopting a suitable feedback between them, the extracted work from the engine can be considered as supplying work for the refrigeration.

\section{Conclusions}
In this work, we provided a modification scheme to present the effect of a generalized measurement channel such as one defined in Eq. (1), on Otto cycle and swap when they are in the refrigerator modes. To this aim, we performed two types of modifications on the familiar Otto cycle refrigerator each of which led to the a different refrigerator. As elucidated, for the first one, the COP depends linearly on the measurement strength parameter so that it goes beyond the Otto limit which in turns is a pure quantum effect. Thus, in comparison to the Otto cycle refrigerator, this type of refrigerator can have potential of interests when one consider it from a finite-time practical realization point of view [33-35]. Since the implementation running time of measurement channel (1) is much smaller than any decoherence or thermalization time scales \cite{Lisboa}, so the total running time of the modified refrigerator seems approximately to be the same as of the Otto cycle one. Thus any finite-time protocol for realization of an Otto cycle refrigerator may be more advantageous for the first type modified refrigerator, which can be the subject of future research.

In the second modification, it was obtained an autonomous refrigerator which does not require to external work source such that the required work for refrigeration process is supplied by a three-stroke measurement-induced  engine whose operation cycle is inside the considered modified Otto cycle. The COP for the obtained refrigerator is the same of the Otto cycle refrigerator as well as its supplying engine efficiency is the same as of the Otto cycle engine. All three strokes of supplying engine depend directly on the intensity of generalized measurement channel so this engine is purely a quantum engine. From the practical point of view, addressing the finite-time realization of such refrigerator would be also interesting. 

Also such modifications were investigated for the two-stroke swap refrigerator. The first modification led to a three-stroke refrigerator whose COP is exactly the same as of the first type modified Otto cycle refrigerator. In the same way, the second modification led to the other three-stroke autonomous refrigerator whose supplying work is provided by a three-stroke engine induced by the measurement process along the same refrigeration cycle.

\newpage
Fig. 1. First type modified Otto cycle refrigerator. The five-stroke refrigeration cycle is obtained through inserting measurement based-stroke before full thermalization of the CM by the cold reservoir in the familiar Otto cycle. In the 4th stroke, the measurement channel, as a quantum resource, absorbs heat $Q(\xi)$ (blue dashed arrow) from the CM. Thus it causes additional heat flow from the cold reservoir into the CM in the 5th stroke, i.e. $Q(\xi)$ (red dashed arrow), so the total heat flow from the CM ($Q_{c}$+$Q(\xi)$) goes beyond the classical cooling limit of the Otto cycle refrigerator. 

\begin{figure}
	\centering
	\includegraphics[width=400 pt]{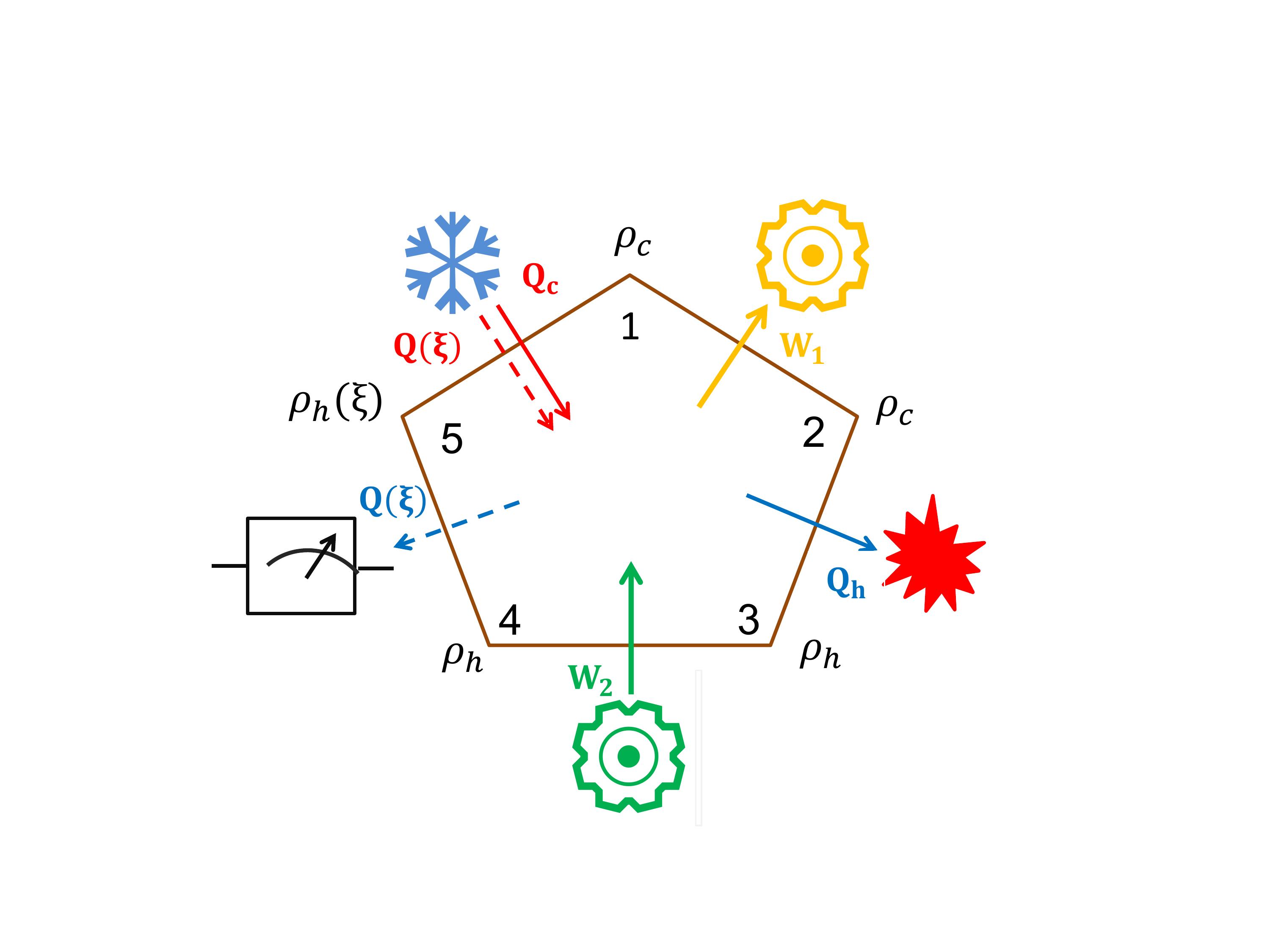}
	\caption{} \label{fig1}
\end{figure}
\newpage
Fig. 2. The COP for the first type modified Otto cycle refrigerator (Eq. (10)), in terms of measurement strength parameter, for some given invested works (Eq. (11)). The parameters $\omega_{h}=10$, $\omega_{c}=2$ and $\beta_c=0.4$ have been fixed. For $\beta_h=0.25$, $\beta_{h}=0.19$, $\beta_{h}=0.16$ and $\beta_{h}=0.14$, the invested works are $W_{in}=1.87$, $W_{in}=1.43$, $W_{in}=1.13$ and $W_{in}=0.89$ respectively and the corresponding COPs are denoted by dotted-dashed black, dashed blue, dotted red and solid green lines respectively. The COP coincides to the Otto limit in $\xi=0$. 
\begin{figure}
	\centering
	\includegraphics[width=300 pt]{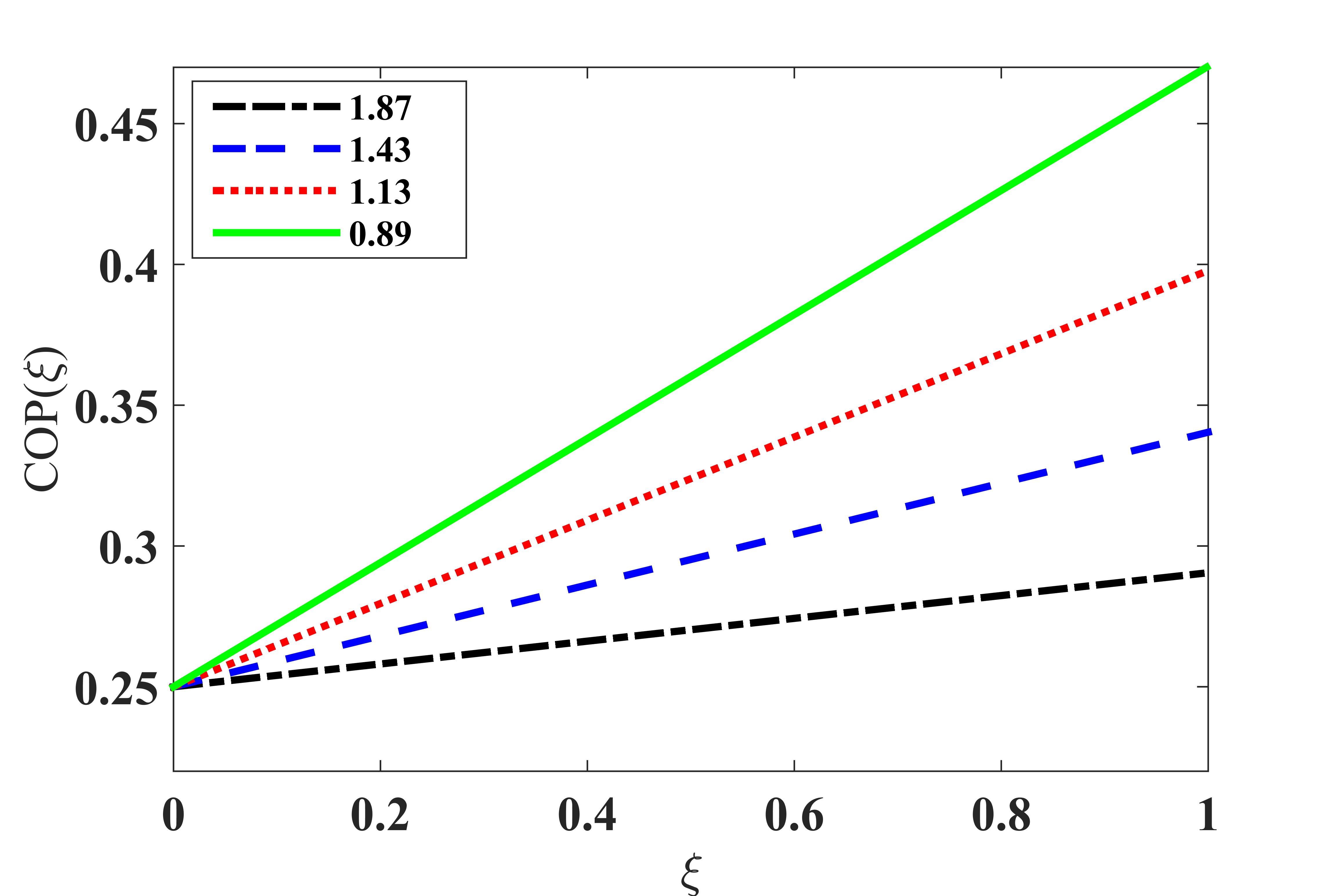}
	\caption{} \label{fig1}
\end{figure}\newpage
Fig. 3. The amount of removal heat from the cold reservoir according to Eq. (8), in terms of measurement strength parameter. The solid green, dotte red, dashed blue and dotted-dashed black lines are corresponding to the invested works (Eq. (11)) $W=0.89$, $W=1.13$, $W=1.43$ and $W=1.89$ respectively. In the strong measurement limit ($\xi\rightarrow1$), the presence of measurement channel makes the most removal heat from the cold reservoir through investing the least amount of work.

\begin{figure}
\centering
\includegraphics[width=300 pt]{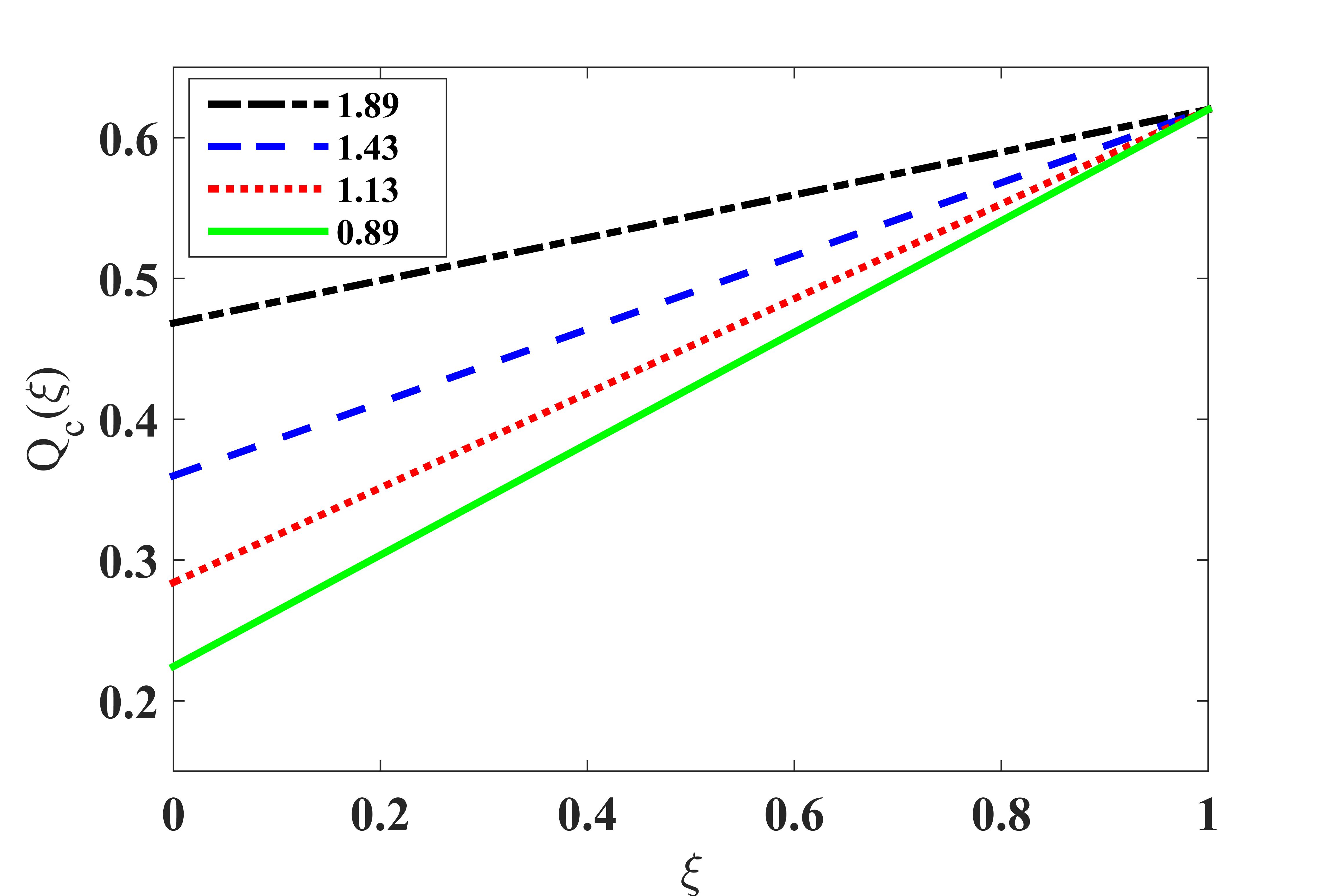}
\caption{} \label{fig1}
\end{figure}
\newpage
Fig. 4. Second type modified Otto cycle refrigerator. The cycle is obtained through inserting the measurement channel after full thermalization of the CM by the cold reservoir in the familiar Otto cycle. It provides not only a refrigeration process in the same way as of Otto cycle refrigerator but also shows simultaneously a three-stroke measurement-induced engine whose efficiency is the same as of Otto cycle engine. The energy exchanges between the two-level system and heat and work sources specified by four solid arrows provide the refrigeration process such that the required work for this purpose is $W_{in}=W_{1}+W_{2}=-(Q_{h}+Q_{c})$. In the same time, the other measurement-based energy exchanges specified by three dashed arrows constitute a quantum engine whose extracted work is $W(\xi)=-(Q'(\xi)+Q(\xi))$. The work $W_{in}$ is provided by an external agent, i.e. $W_{ex}(\xi)$, as well as by the measurement-induced engine, i.e. $W(\xi)$ (through suitable feedback) such that $W_{in}=W_{ex}(\xi)-W(\xi)$ (see Eq. (24)). In the limit $\xi\rightarrow \xi_{c}$, $W_{ex}(\xi_{c})=0$, i.e. the refrigeration process has no need to work supplying by an external agent. In this situation the modified Otto cycle constitutes an autonomous refrigerator whose invested work is only provided by the work extracted from the measurement-induced engine, i.e. $W_{in}=-W(\xi_{c})$.

\begin{figure}
	\centering
	\includegraphics[width=400 pt]{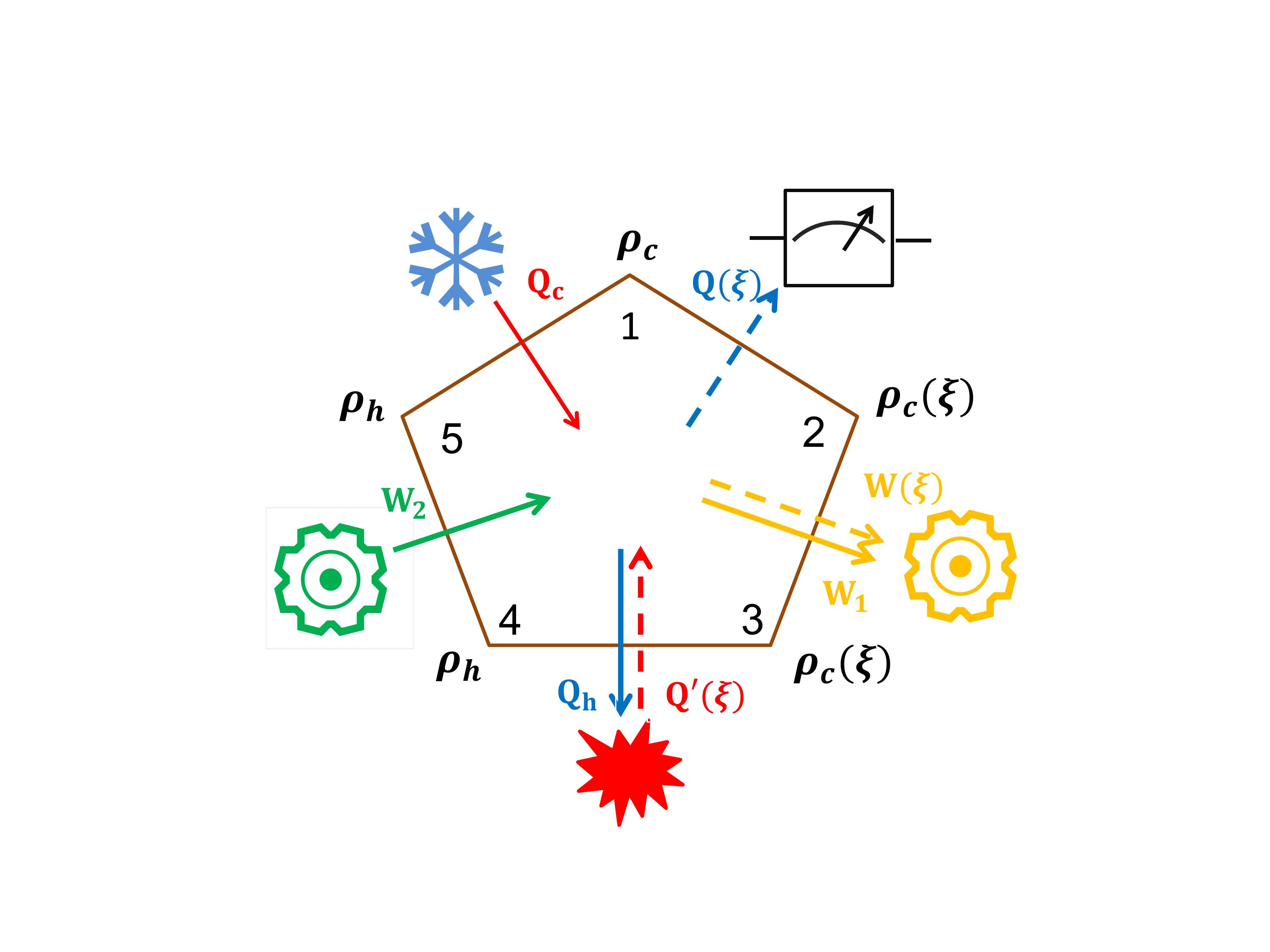}
	\caption{} \label{fig1}
\end{figure}
\newpage
Fig. 5. First type modified swap refrigerator. (a) At the beginning, each subsystem has been isolated from its own reservoir then the state of bipartite system is $\rho_{h}\otimes\rho_{c}$. (b) The 1st stroke is related to application of swap operator on the bipartite system which makes the cold qubit becomes colder and the hot one becomes hotter. This requires performing the work $W_{in}$ (green solid arrow) on the CM, supplied by an external agent. (c) The effect of measurement (2nd stroke) on the colder qubit allows flow of quantum heat $Q(\xi)$ (blue dashed arrow) from CM into the measurement apparatus. (d) In the 3rd stroke, each subsystem is brought into contact with its own reservoir so the CM absorbs heats $(Q_{c}+Q(\xi))$ (red solid and dashed arrows respectively) from the cold reservoir and loses heat $Q_{h}$ into the hot reservoir. The output of this refrigerator is the same as of the first type modified Otto cycle refrigerator.  

\begin{figure}
	\centering
(a)\qquad\qquad \quad\qquad\quad\quad\qquad\qquad\quad\qquad\quad\qquad (b)\\{
	\includegraphics[width=231 pt]{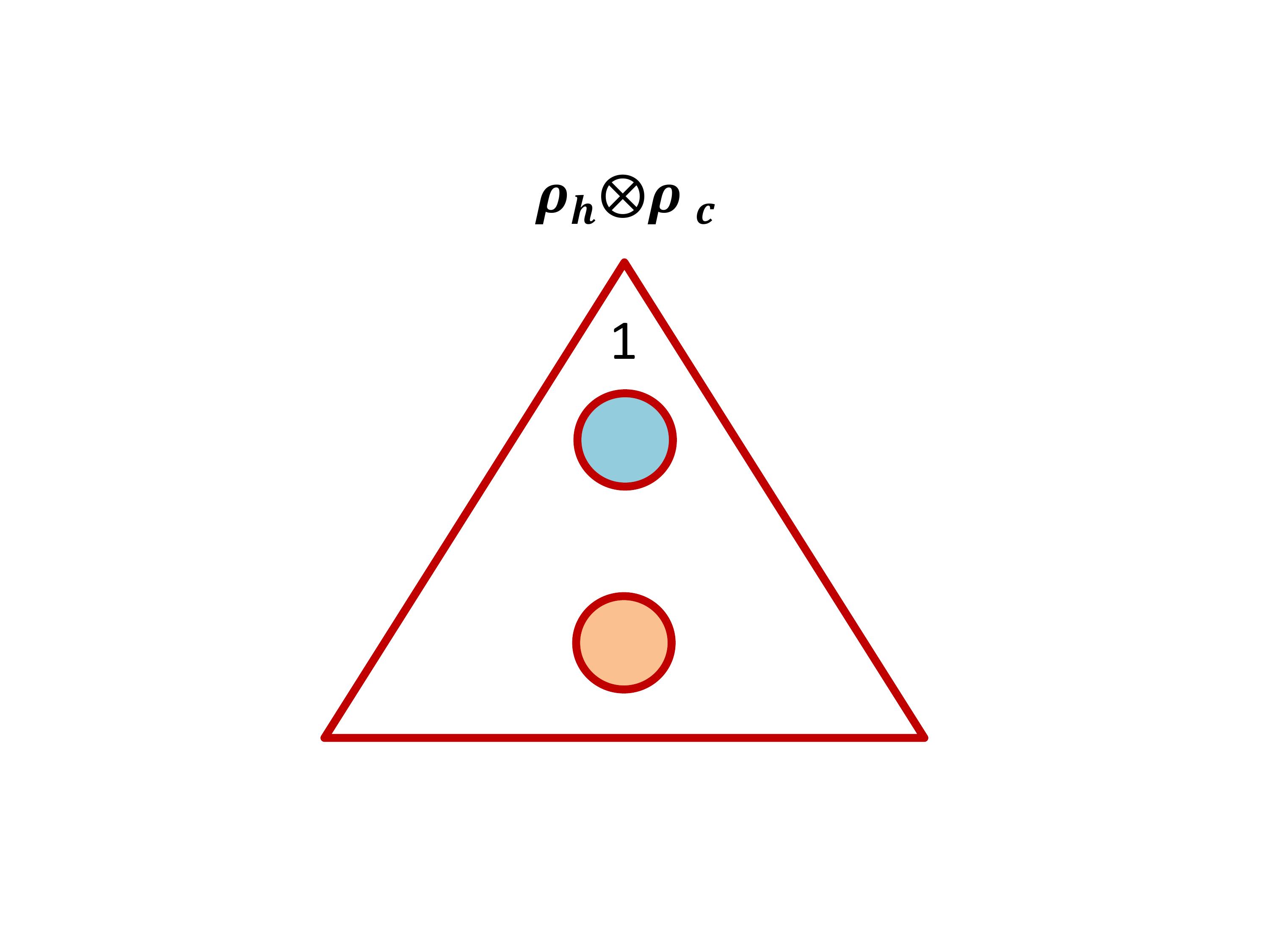}
	\label{fig:first_sub}
}{
	\includegraphics[width=231 pt]{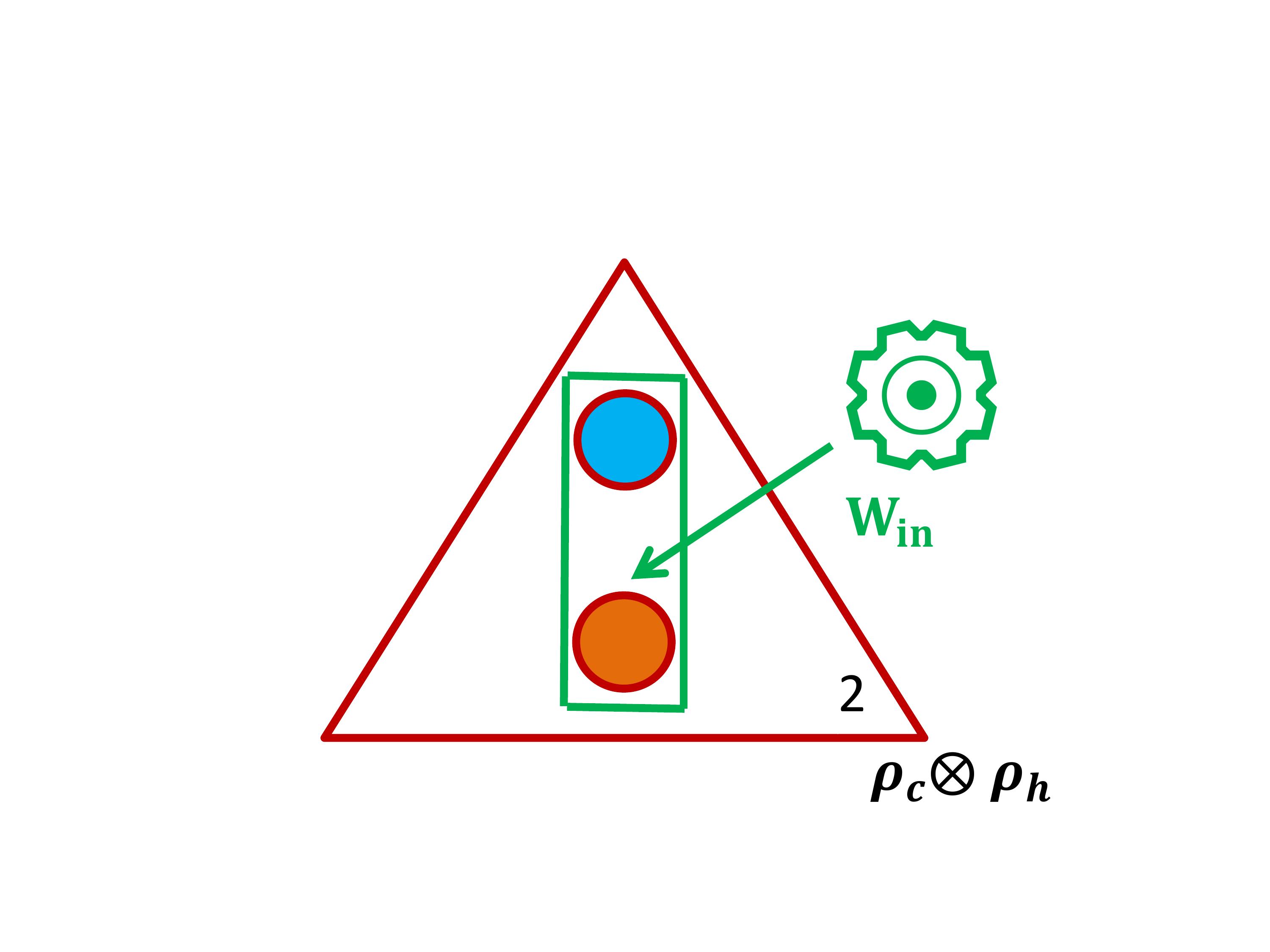}
	\label{fig:second_sub}
}(c)\qquad\qquad \quad\qquad\quad\quad\qquad\qquad\quad\qquad\quad\qquad (d)\\{
\includegraphics[width=231 pt]{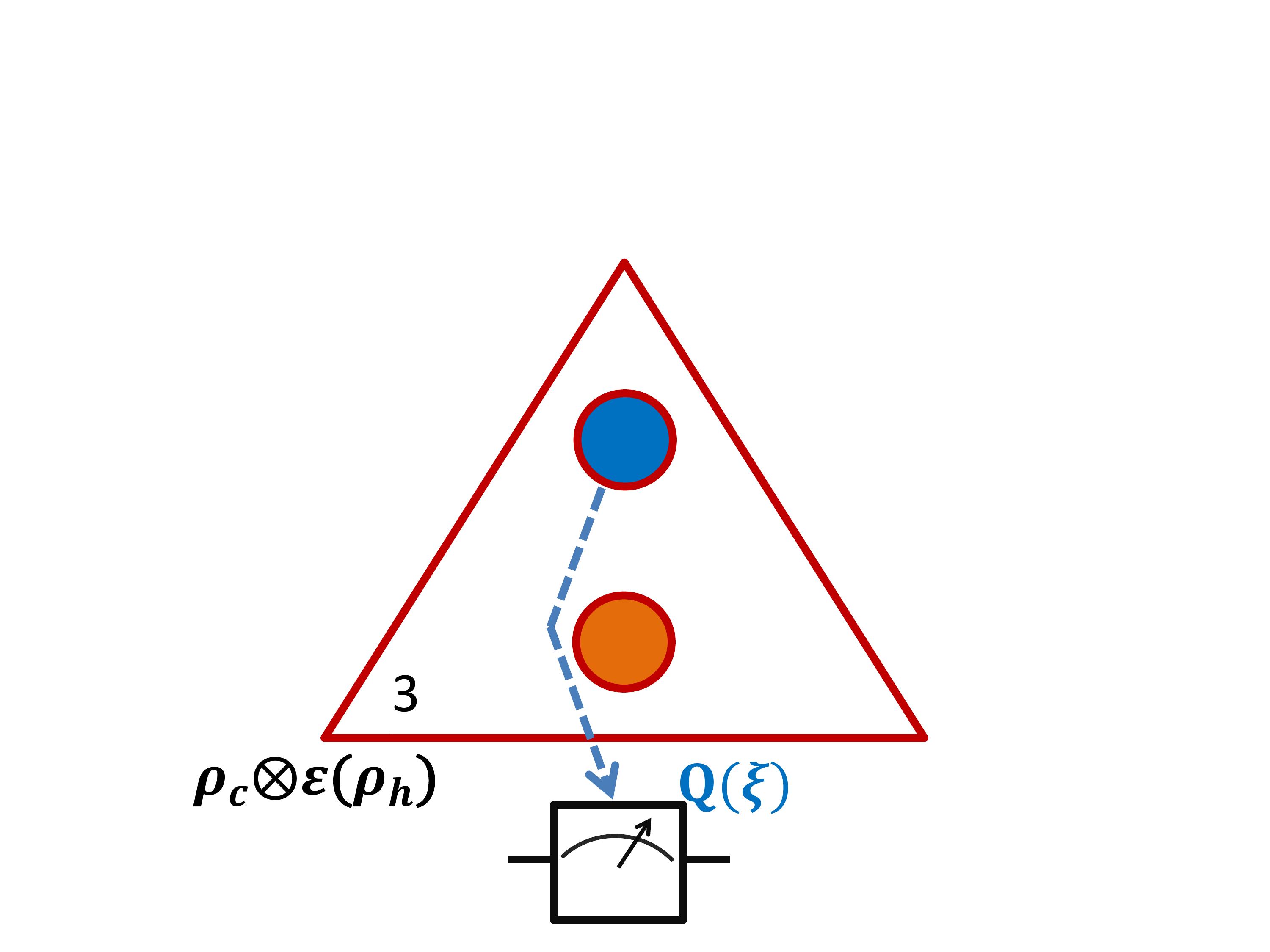}
\label{fig:first_sub}
}{
\includegraphics[width=231 pt]{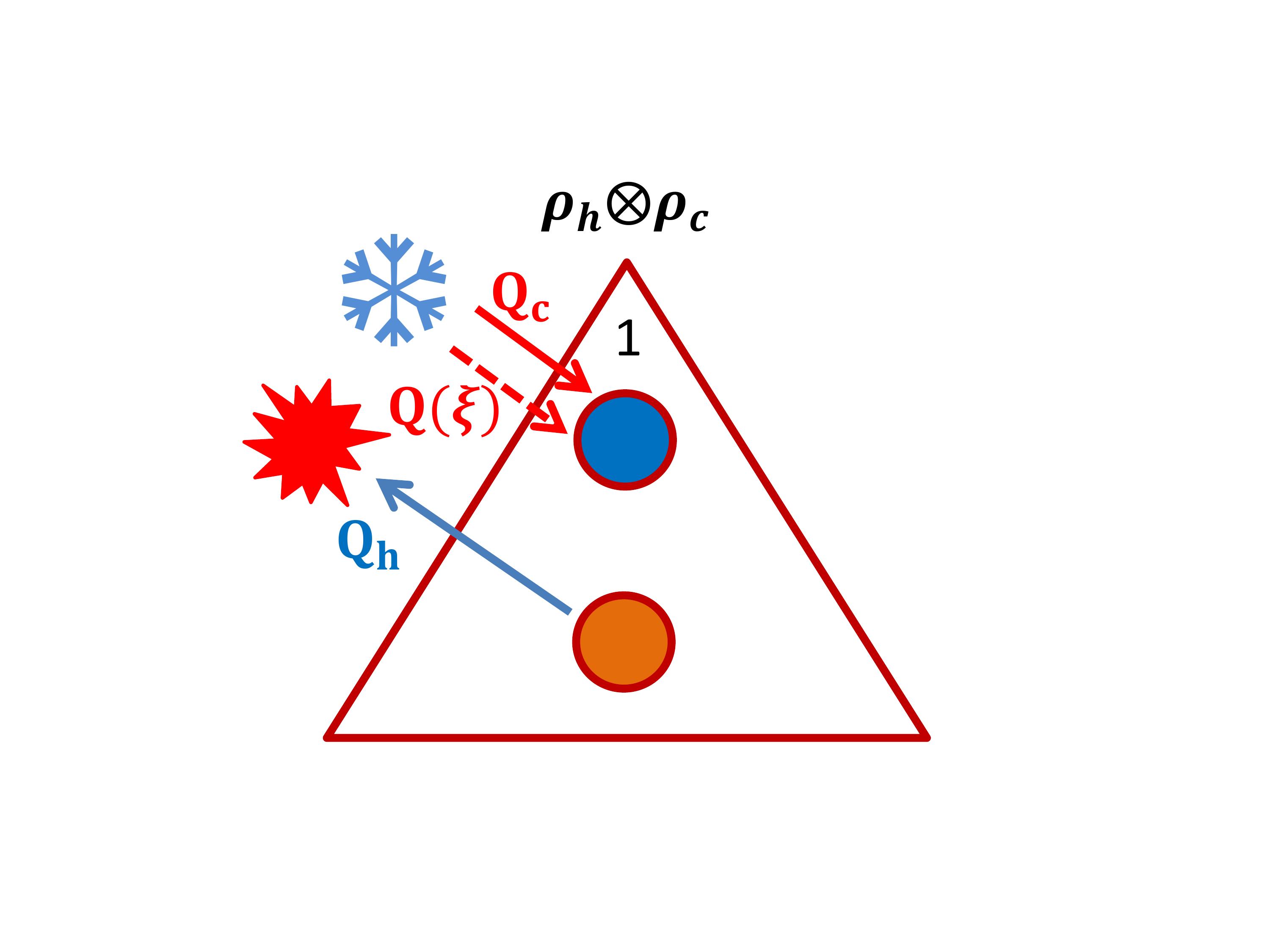}
\label{fig:second_sub}
}
	\caption{} \label{fig1}
\end{figure}
\newpage
Fig. 6. Second type modified swap refrigerator. (a) At the beginning, each subsystem has been isolated from its own reservoir so they are described by a tensor product state $\rho_{h}\otimes\rho_{c}$. (b) The subsystem with state $\rho_{c}$ is affected by the measurement (1st stroke), so the quantum heat $Q(\xi)$ (blue dashed arrow) is transfered from the mentioned subsystem into the measurement apparatus. (c) By the swap operation (2nd stroke), the required work for the refrigeration, i.e. $W_{in}$ (green solid arrow), is supplied by an external agent, i.e. $W_{ex}(\xi)$ as well as by the measurement-induced engine whose extracted work is $W(\xi)$ (red dashed arrow) such that $W_{in}=W_{ex}(\xi)-W(\xi)$ (see Eq. (34)). At the end of this stroke, the cold qubit becomes colder and the hot one becomes hotter. (d) The 3rd stroke is provided by bringing each subsystem into contact with its own reservoir. During this stroke, the heat $Q_{c}$ (red solid arrow) is removed from the cold reservoir and heat $Q_{h}$ (blue solid arrow) is transfered from the CM into the hot reservoir, whereas the heat $Q'(\xi)$ (red dashed arrow) is absorbed from the hot reservoir by the CM. Clearly $W_{in}=-(Q_{h}+Q_{c})$ and $W(\xi)=-(Q'(\xi)+Q(\xi))$. In the limit $\xi\rightarrow\xi_{c}$ (see Sec. III), the modified swap provides an autonomous refrigerator where the required work for the refrigeration is only supplied by the measurement-induced engine. The outputs of this cycle are the same as of second type modified Otto cycle. 
\begin{figure}
	\centering
	(a)\qquad\qquad \quad\qquad\quad\quad\qquad\qquad\quad\qquad\quad\qquad (b)\\{
		\includegraphics[width=231 pt]{Fig5}
		\label{fig:first_sub}
	}{
		\includegraphics[width=231 pt]{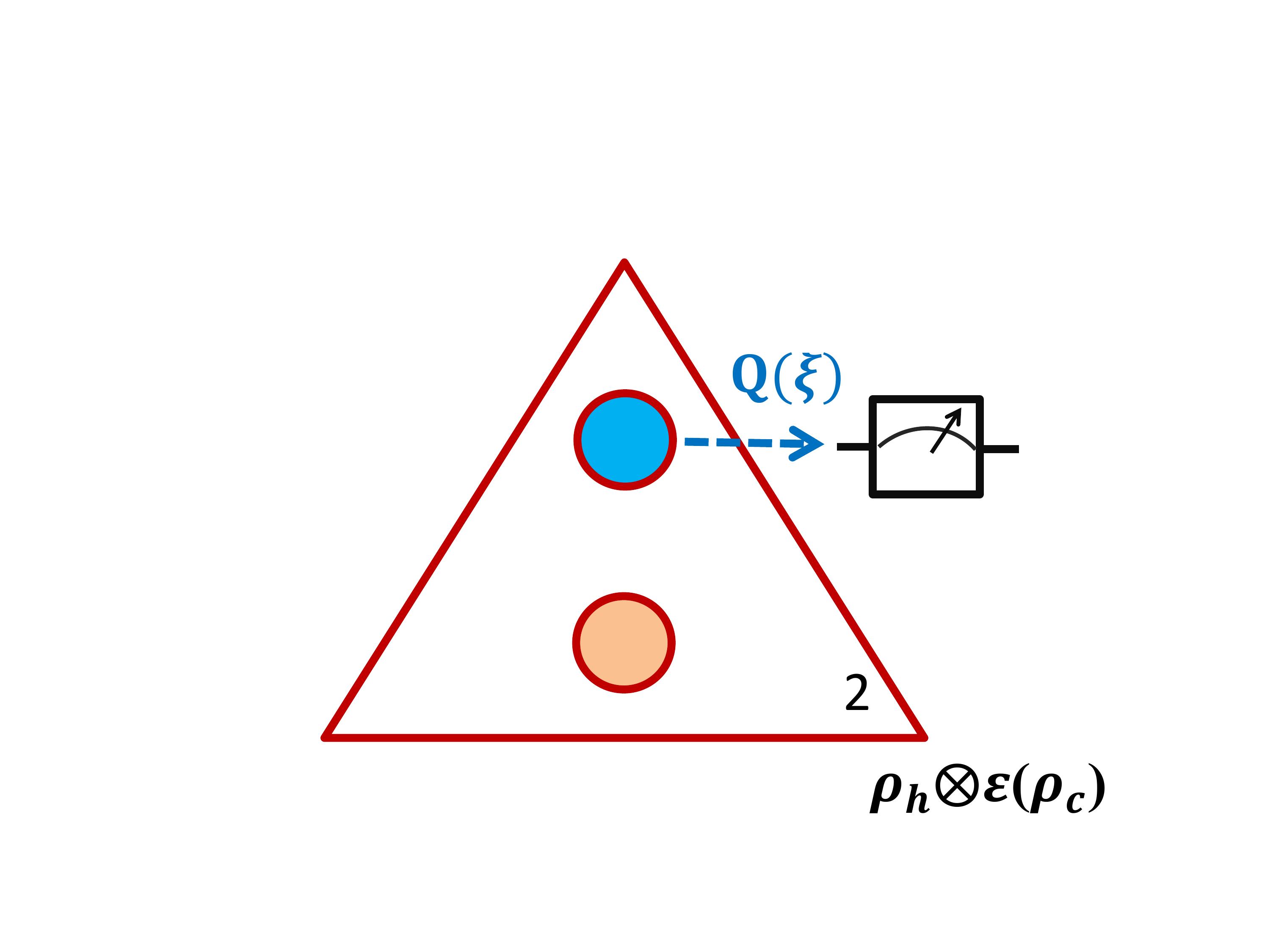}
		\label{fig:second_sub}
	}(c)\qquad\qquad \quad\qquad\quad\quad\qquad\qquad\quad\qquad\quad\qquad (d)\\{
		\includegraphics[width=231 pt]{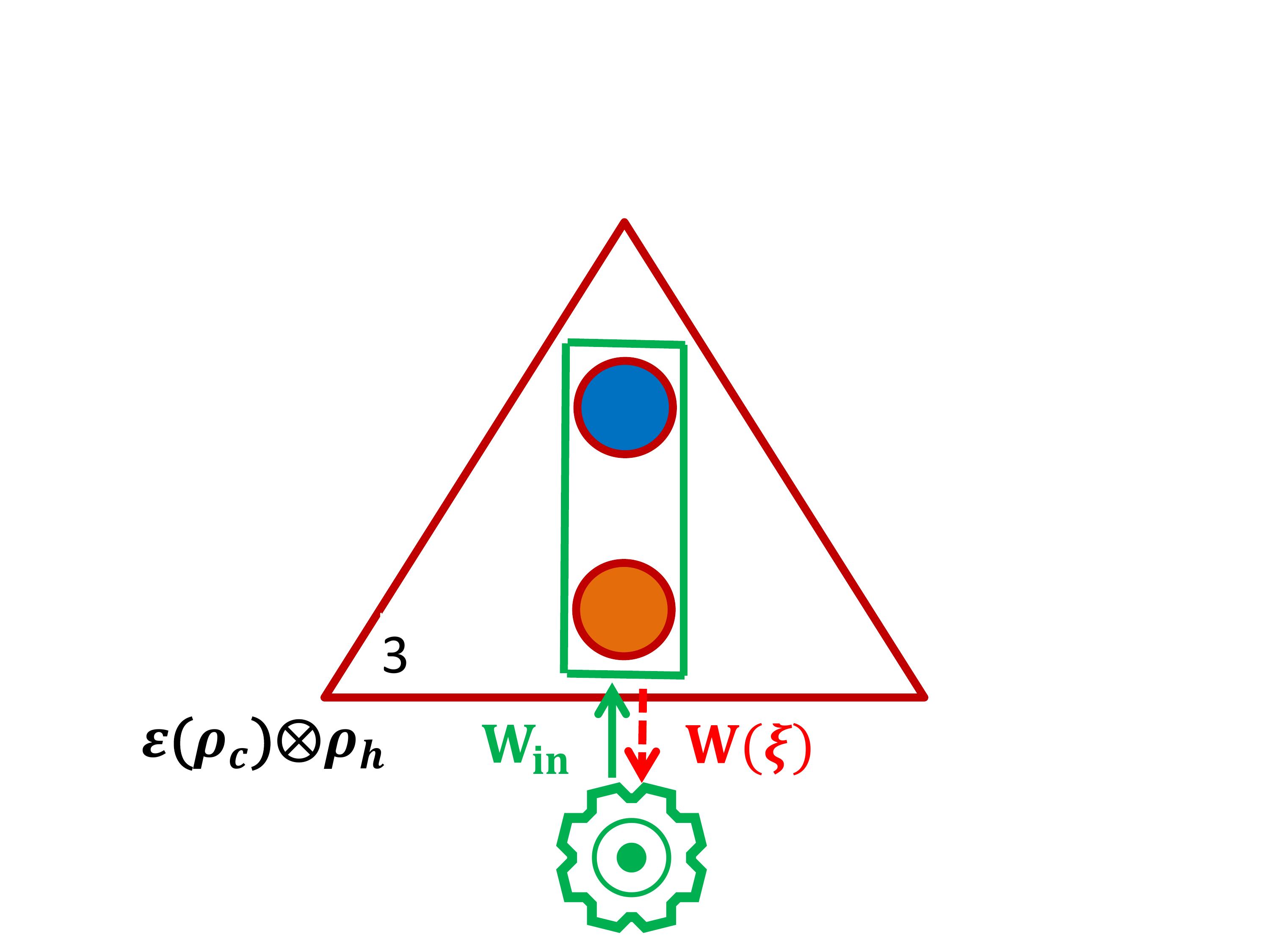}
		\label{fig:first_sub}
	}{
		\includegraphics[width=231 pt]{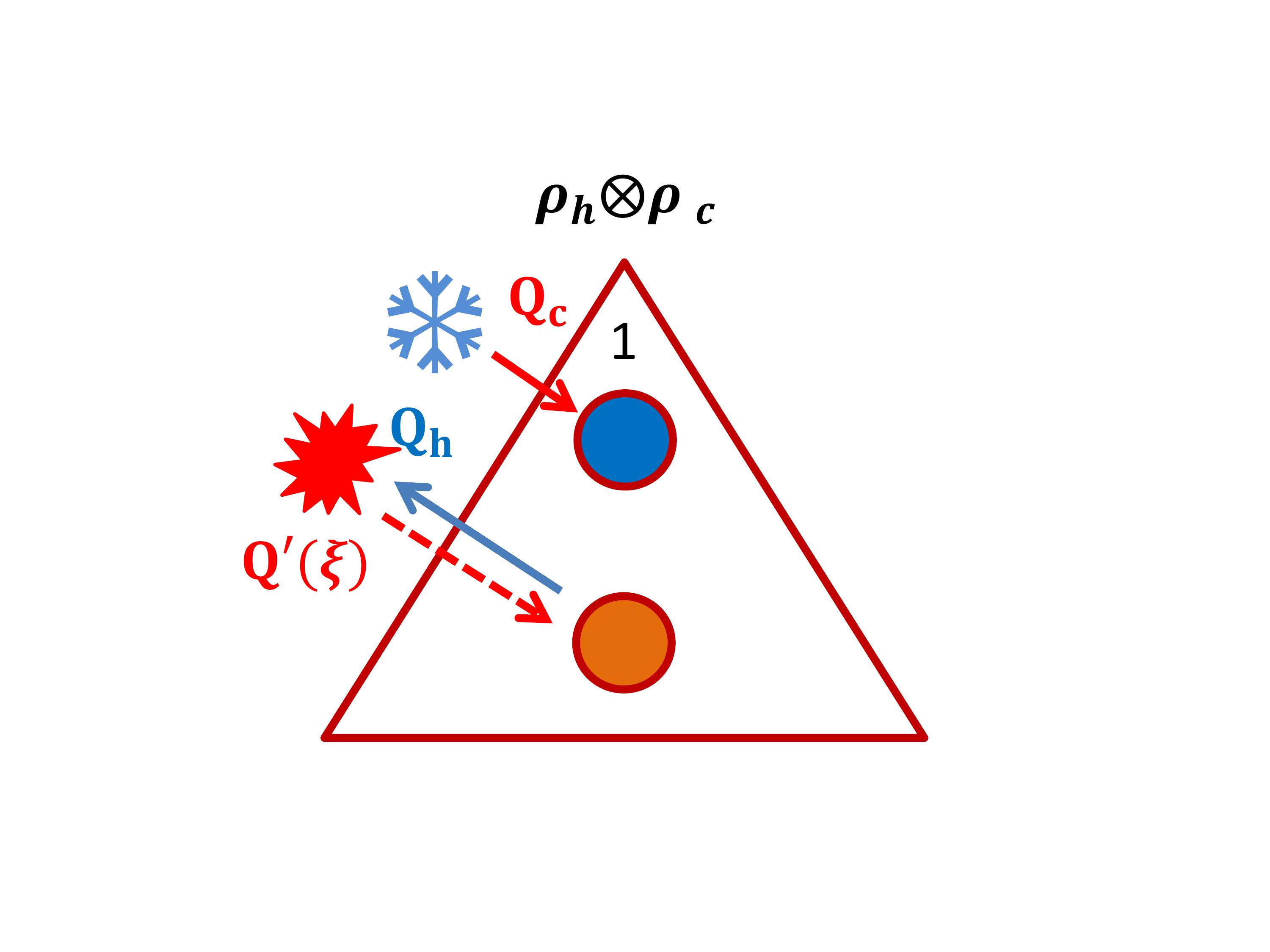}
		\label{fig:second_sub}
	}
	\caption{} \label{fig1}
\end{figure}
\end{document}